\documentstyle[preprint,aps]{revtex}
\widetext
\draft
\tighten

\begin{document}

\title{Does the electromagnetic field of an accelerated charge satisfy
Maxwell equations?}

\bigskip

\author{\bf Andrew E. Chubykalo and Stoyan J. Vlaev}

\address {Escuela de F\'{\i}sica, Universidad Aut\'onoma de Zacatecas \\
Apartado Postal C-580\, Zacatecas 98068, ZAC., M\'exico}

\date{\today}

\maketitle


\baselineskip 7mm

\begin{abstract}
We considered the electromagnetic field of a charge moving with
a constant acceleration along an axis. We found that this field
obtained from the Li\'enard-Wiechert potentials does not satisfy Maxwell
equations.
\end{abstract}


$$$$
{\bf I. INTRODUCTION}

It is well-known that the electromagnetic field created by an arbitrarily
moving charge
   \begin{equation}
   {\bf E}({\bf r},t)=q\left\{\frac{({\bf R}-R\frac{{\bf
         V}}{c})(1-\frac{V^{2}}{c^{2}})}{(R-{\bf R}\frac{{\bf
   V}}{c})^{3}}\right\}_{t_0}+q\left\{\frac{[{\bf R}\times[({\bf
   R}-R\frac{{\bf V}}{c})\times\frac{{\bf{\dot{V}}}}{c^{2}}]]}{(R-{\bf
   R}\frac{{\bf V}}{c})^{3}}\right\}_{t_0},
   \end{equation}
\begin{equation}
{\bf B}({\bf r},t)=\left\{\left[\frac{{\bf R}}{R}\times{\bf
E}\right]\right\}_{t_0}
\end{equation}

was obtained directly from Li\'enard-Wiechert
potentials$^1$:

\begin{equation}
\varphi({\bf r},t)=\left\{\frac{q}{\left(R-{\bf R}\frac{{\bf
   V}}{c}\right)}\right\}_{t_0},\qquad
{\bf A}({\bf r},t)=\left\{\frac{q{\bf V}}{c\left(R-{\bf R}\frac{{\bf
   V}}{c}\right)}\right\}_{t_0}.
\end{equation}

  Usually, the first terms of the right-hand sides ({\it rhs}) of (1) and
 (2) are called ``velocity fields" and the second ones are called
 ``acceleration fields".

It was recently
claimed by E.Comay$^2$ that ``...  {\it Acceleration
fields by themselves do not satisfy Maxwell's equations$^3$. Only the sum
of velocity fields and acceleration fields satisfies Maxwell's
equations}." We wish to argue that this sum {\it does not satisfy}
Maxwell's equations:  \begin{eqnarray} && \nabla\cdot{\bf
E}=4\pi\varrho,\\ && \nabla\cdot{\bf B}=0,\\ && \nabla\times{\bf
H}=\frac{4\pi}{c}{\bf j}+\frac{1}{c}\frac{\partial {\bf E}}{\partial t},\\
&& \nabla\times{\bf E}=-\frac{1}{c}\frac{\partial {\bf B}}{\partial t}.
\end{eqnarray}

First, let us recall the usual way of deriving the formulas (1), (2) for
the electric ({\bf E}) and magnetic ({\bf B}) fields$^1$.

To obtain the values
of $\varphi$, {\bf A} (see Eq.(3)) and {\bf E}, {\bf B} (see Eqs. (1) and
(2)) at the instant $t$ one has to take the values of {\bf V}, $\dot{\bf
 V}$ and {\bf R} at instant $t_0$.  Here $t_0=t-\tau$, $\tau$ is the so
 called ``retarded time", ${\bf R}$ is the vector connecting the site
 ${\bf r}_0(x_0,y_0,z_0)$ of the charge $q$ at instant $t_0$ with the
 point of observation {\bf r}($x,y,z$). The instant $t_0$ is determined
 from the condition (see Eq.(63.1) of Ref.1):  \begin{equation}
 t_0=t-\tau=t-\frac{R(t_0)}{c}.  \end{equation}

 The  {\it rhs} of (3) contain functions of $t_0$,
 which, in turn, depends on $x,y,z,t$:

 \begin{equation}
t_0=f(x,y,z,t).
 \end{equation}

To calculate the fields {\bf E} and {\bf B} one has to substitute
$\varphi$ and {\bf A} from (3) in the following expressions$^1$:
\begin{equation}
{\bf E}=-\nabla\varphi-\frac{1}{c}\frac{\partial{\bf A}}{\partial
t},\qquad {\bf B}=[\nabla\times{\bf A}].
\end{equation}

Substituting $\varphi, A_x, A_y, A_z$ given by (3) in Eq.(10), one ought
to calculate $\partial\{\}/\partial t$ and
$\partial\{\}/\partial x_i$ ($x_i$ are $x,y,z$) using the
following scheme:

\medskip

\begin{equation}
\left.
{\Large
\begin{array}{c}
{
\frac{\partial\varphi}{\partial x_i}=
\frac{\partial\varphi}{\partial t_0}\frac{\partial t_0}{\partial x_i},
}\\
{
 }\\
{\frac{\partial{\bf A}}{\partial t}=
\frac{\partial{\bf A}}{\partial t_0}\frac{\partial t_0}{\partial t},
}\\
{
 }\\
{\frac{\partial A_k}{\partial x_i}=
\frac{\partial A_k}{\partial t_0}\frac{\partial t_0}{\partial x_i},
}
\end{array}}
\right\}
\end{equation}
\medskip
and as a result one obtains the formulas (1) and (2).

In the next section we will consider a charge moving with a constant
acceleration along the $X$ axis and we will show that the Eq.(7) is not
satisfied if one substitutes ${\bf E}$ and ${\bf B}$ from Eqs.(1) and (2)
in Eq.(7). To verify this we have to find the derivatives of
$x-,y-,z-$components of the fields ${\bf E}$ and ${\bf B}$ with respect to
the time $t$ and the coordinates $x,y,z$. The functions ${\bf E}$ and
${\bf B}$ depend on $x,y,z,t$ through $t_0$ from the conditions (8)-(9).
In other words, we will show that {\it these fields ${\bf E}$ and ${\bf
B}$ do not satisfy the Maxwell equations if the differentiation rules that
were applied to $\varphi$ and ${\bf A}$ (to obtain ${\bf E}$ and ${\bf
B}$) are applied identically to ${\bf E}$ and ${\bf B}$}.

\clearpage

$$$$
{\bf II. DOES THE ELECTROMAGNETIC FIELD OF A CHARGE MOVING WITH A CONSTANT
ACCELERATION SATISFY MAXWELL EQUATIONS?}

Let us consider a charge $q$ moving with a constant acceleration
along the $X$ axis. In this case its velocity and acceleration have only
$x$-components, respectively ${\bf V}(V,0,0)$ and ${\bf a}(a,0,0)$.
Now we will rewrite the Eqs. (1) and (2) by components:

\begin{equation}
E_x(x,y,z,t)=q\left\{\frac{(V^2-c^2)[RV-c(x-x_0)]}{\left[(cR-
V(x-x_0)\right]^3}\right\}_{t_0}   + q\left\{
\frac{ac[(x-x_0)^2-R^2]}{\left[(cR-V(x-x_0)\right]^3}\right\}_{t_0},
\end{equation}

\begin{equation}
E_y(x,y,z,t)=-q\left\{\frac{c(V^2-c^2)(y-y_0)}{\left[(cR-V(x-
x_0)\right]^3}\right\}_{t_0} + q\left\{
\frac{ac(x-x_0)(y-y_0)}{\left[(cR-V(x-x_0)\right]^3}\right\}_{t_0},
\end{equation}

\begin{equation}
E_z(x,y,z,t)=-q\left\{\frac{c(V^2-c^2)(z-z_0)}{\left[(cR-V(x-
x_0)\right]^3}\right\}_{t_0} + q\left\{
\frac{ac(x-x_0)(z-z_0)}{\left[(cR-V(x-x_0)\right]^3}\right\}_{t_0},
\end{equation}

\begin{equation}
B_x(x,y,z,t)=0,
\end{equation}

\begin{equation}
B_y(x,y,z,t)=q\left\{\frac{V(V^2-c^2)(z-z_0)}{\left[(cR-V(x-
x_0)\right]^3}\right\}_{t_0} - q\left\{
\frac{acR(z-z_0)}{\left[(cR-V(x-x_0)\right]^3}\right\}_{t_0},
\end{equation}

\begin{equation}
B_z(x,y,z,t)=-q\left\{\frac{V(V^2-c^2)(y-y_0)}{\left[(cR-
V(x-x_0)\right]^3}\right\}_{t_0} + q\left\{
\frac{acR(y-y_0)}{\left[(cR-V(x-x_0)\right]^3}\right\}_{t_0},
\end{equation}

\medskip

Obviously, these components are functions of $x,y,z,t$  through $t_0$ from
the conditions (8)-(9). This means that when substituting the field
components given by Eqs.(12)-(17) in the Maxwell equations (4)-(7), we
still have to use the differentiation rules as in (11):

\medskip

\begin{equation}
\left.
{\Large
\begin{array}{c}
{\frac{\partial E\{{\rm or}\,B\}_k}{\partial t}=
\frac{\partial E\{{\rm or}\, B\}_k}{\partial t_0}\frac{\partial
t_0}{\partial t}, }\\
{
 } \\
{\frac{\partial E\{{\rm or}\,B\}_k}{\partial x_i}=
\frac{\partial E\{{\rm or}\, B\}_k}{\partial t_0}\frac{\partial
t_0}{\partial x_i},
}
\end{array}}
\right\}
\end{equation}
\medskip
where $k$ and $x_i$ are $x,y,z$.

To calculate $\partial t_0/\partial t$ and $\partial t_0/\partial x_i$ one
ought to use differentiation rules for implicit functions:
\begin{equation}
 \frac{\partial t_0}{\partial t}=-\frac{{\partial F}/{\partial t}}
 {{\partial F}/{\partial t_0}};\qquad \frac{\partial t_0}{\partial
 x_i}=-\frac{{\partial F}/{\partial x_i}} {{\partial F}/{\partial t_0}},
 \end{equation}
where
\begin{equation}
F(x,y,z,t,t_0)=t-t_0-\frac{R}{c}=0,\qquad
R=\left(\sum_{i}[(x_i-x_{0i}(t_0)]^2\right)^{1/2}.
\end{equation}
In this case one obtains:
\begin{equation}
\frac{\partial t_0}{\partial t}=\frac{R}{R-(x-x_0)V/c}
\qquad
{\rm and}\qquad
\frac{\partial t_0}{\partial x_i}=-
\frac{x_i-x_{0i}}{c[R-(x-x_0)V/c]}.
\end{equation}
Remember that we are considering the case with ${\bf V}=(V,0,0)$ here.
There is a different way to calculate the derivatives (19)$^4$,
which gives the expressions (21).

Let us rewrite Eq.(7) by components taking into account the rules (18) and
Eq.(15):

\medskip

\begin{equation}
\frac{\partial E_z}{\partial t_0}\frac{\partial t_0}{\partial y}
-\frac{\partial
E_y}{\partial t_0}\frac{\partial t_0}{\partial z}=0,
\end{equation}

\begin{equation}
\frac{\partial E_x}{\partial t_0}\frac{\partial t_0}{\partial z}
-\frac{\partial
E_z}{\partial t_0}\frac{\partial t_0}{\partial x}+
\frac{1}{c}\frac{\partial B_y}{\partial t_0}\frac{\partial t_0}
{\partial t}
=0,
\end{equation}

\begin{equation}
\frac{\partial E_y}{\partial t_0}\frac{\partial t_0}{\partial x}
-\frac{\partial
E_x}{\partial t_0}\frac{\partial t_0}{\partial y}+
\frac{1}{c}\frac{\partial B_z}{\partial t_0}\frac{\partial t_0}
{\partial t}
=0.
\end{equation}

\medskip

In order to calculate the derivatives $\partial E({\rm or}\,B)_k/\partial
t_0$ we have to know the values of the expressions  $\partial V/\partial
t_0$, $\partial x_0/\partial t_0$ and $\partial R/\partial t_0$. Note that
to find $\partial\varphi/\partial t_0$ from (3) one uses the following
values of these expressions$^5$:  \begin{equation} \frac{\partial
R}{\partial t_0}=-c,\qquad\frac{\partial{\bf R}}{\partial
t_0}=-\frac{\partial{\bf r}_0}{\partial t_0}=-{\bf V}(t_0)\qquad {\rm and}
\qquad\frac{\partial{\bf V}}{\partial t_0}=\dot{\bf V}.  \end{equation}
So, in our case we have to use \begin{equation} \frac{\partial R}{\partial
t_0}=-c,\qquad \frac{\partial x_0}{\partial t_0}=V\qquad{\rm and}\qquad
\frac{\partial V}{\partial t_0}=a.  \end{equation}

Now, using Eqs. (21) and (26), we want to verify the validity of
Eqs.(22)-(24).  The result of the verification is as follows$^6$:

\begin{equation}
\frac{\partial E_z}{\partial t_0}\frac{\partial t_0}{\partial y}
-\frac{\partial
E_y}{\partial t_0}\frac{\partial t_0}{\partial z}=0,
\end{equation}

\begin{equation}
\frac{\partial E_x}{\partial t_0}\frac{\partial t_0}{\partial z}
-\frac{\partial
E_z}{\partial t_0}\frac{\partial t_0}{\partial x}+
\frac{1}{c}\frac{\partial B_y}{\partial t_0}\frac{\partial t_0}
{\partial t}
=-\frac{ac(z-z_0)}{[cR-V(x-x_0)]^3},
\end{equation}

\begin{equation}
\frac{\partial E_y}{\partial t_0}\frac{\partial t_0}{\partial x}
-\frac{\partial
E_x}{\partial t_0}\frac{\partial t_0}{\partial y}+
\frac{1}{c}\frac{\partial B_z}{\partial t_0}\frac{\partial t_0}
{\partial t}
=\frac{ac(y-y_0)}{[cR-V(x-x_0)]^3}.
\end{equation}

The
verification shows that Eq.(22) is valid. But instead of Eq.(23) and
Eq.(24) we have Eq.(28) and Eq.(29) respectively.

For the present we refrain from any comment regarding this result.
However, we would like to cite the following phrase from the recent
work$^7$:  ``Maxwell equations may not be an adequate description of
nature".

$$$$
{\bf APPENDIX}

To obtain Eqs. (1) and (2), let us rewrite Eqs.(10) taking into account
Eqs.(11)$^8$:

\begin{equation}
{\bf E}=-\nabla\varphi-\frac{1}{c}\frac{\partial{\bf A}}{\partial
t}=-\frac{\partial\varphi}{\partial t_0}\nabla
t_0-\frac{1}{c}\frac{\partial {\bf A}}{\partial t_0}\frac{\partial
t_0}{\partial t},
\end{equation}
\begin{equation}
{\bf B}=[\nabla\times{\bf A}]=\left[\nabla t_0\times\frac{\partial {\bf
A}}{\partial t_0}\right].
\end{equation}

From Eqs.(3) we obtain:
\begin{equation}
\frac{\partial\varphi}{\partial t_0}=-\frac{q}{(R-{\bf R}
\mbox{\boldmath$\beta$})^2}\left(\frac{\partial R}{\partial
t_0}-\frac{\partial{\bf R}}{\partial t_0}\mbox{\boldmath$\beta$}-{\bf
R}\frac{\partial\mbox{\boldmath$\beta$}}{\partial t_0}\right),
\end{equation}
where $\mbox{\boldmath$\beta$}={\bf V}/c$. Hence, taking into account
Eqs.(25), we have (after some algebraic simplification):
\begin{equation}
\frac{\partial\varphi}{\partial t_0}=\frac{qc(1-\beta^2+ {\bf
R}\dot{\mbox{\boldmath$\beta$}}/c)}{(R-{\bf R}
\mbox{\boldmath$\beta$})^2}.
\end{equation}
In  turn
\begin{equation}
\frac{\partial {\bf A}}{\partial t_0}=\frac{\partial\varphi}{\partial t_0}
\mbox{\boldmath$\beta$}+\varphi\dot{\mbox{\boldmath$\beta$}}.
\end{equation}
Putting $\varphi$ from Eqs.(3), Eq.(33) and Eq.(34) together, we obtain
(after simplification):

\begin{equation}
\frac{\partial {\bf A}}{\partial
t_0}=qc\,\,\frac{\mbox{\boldmath$\beta$}(1- \beta^2+ {\bf
R}\dot{\mbox{\boldmath$\beta$}}/c)+
(\dot{\mbox{\boldmath$\beta$}}/c)(R-{\bf R}
\mbox{\boldmath$\beta$})}{(R-{\bf R}
\mbox{\boldmath$\beta$})^2}.
\end{equation}
Finally, substituting Eqs. (21)$^9$, (33) and (35) in Eq.(30) we obtain:
\begin{eqnarray}
{\bf E} &=& \frac{qc(1-\beta^2+ {\bf
R}\dot{\mbox{\boldmath$\beta$}}/c)}{(R-{\bf R}
\mbox{\boldmath$\beta$})^2}\left(-\frac{{\bf R}}{c(R-{\bf
R}\mbox{\boldmath$\beta$})}\right)- \nonumber\\
&& \nonumber\\
&& -\,\,q\,\,\frac{\mbox{\boldmath$\beta$}(1- \beta^2+ {\bf
R}\dot{\mbox{\boldmath$\beta$}}/c)+
(\dot{\mbox{\boldmath$\beta$}}/c)(R-{\bf R}
\mbox{\boldmath$\beta$})}{(R-{\bf R}
\mbox{\boldmath$\beta$})^2}\left(\frac{R}{R-{\bf
R}\mbox{\boldmath$\beta$}}\right)= \nonumber\\
&& \nonumber\\
&& = q\,\,\frac{{\bf R}(1-\beta^2+ {\bf
R}\dot{\mbox{\boldmath$\beta$}}/c)-R\mbox{\boldmath$\beta$} (1-\beta^2+
{\bf
R}\dot{\mbox{\boldmath$\beta$}}/c)-(R\dot{\mbox{\boldmath$\beta$}}/c)(R-
{\bf R}\mbox{\boldmath$\beta$})}{(R-{\bf R}\mbox{\boldmath$\beta$})^3}.
\end{eqnarray}

Grouping together all terms with acceleration together, one can reduce
this expression to \begin{equation} {\bf E}=q\,\frac{({\bf R}-R\frac{{\bf
         V}}{c})(1-\frac{V^{2}}{c^{2}})}{(R-{\bf R}\frac{{\bf
   V}}{c})^{3}}+q\,\frac{({\bf
R}\dot{\mbox{\boldmath$\beta$}}/c)({\bf R}-R\mbox{\boldmath$\beta$})-
(R\dot{\mbox{\boldmath$\beta$}}/c)(R- {\bf R}\mbox{\boldmath$\beta$})}{
(R-{\bf R}\mbox{\boldmath$\beta$})^3}.
\end{equation}
Now, using the formula of double vectorial product$^{10}$, it is not worth
reducing the numerator of the second term of Eq.(37) to  $[{\bf
R}\times[({\bf R}-R\mbox{\boldmath$\beta$})\times
\dot{\mbox{\boldmath$\beta$}}/c]]$. As a result we have Eq.(1).

Analogically, substituting Eqs. (21) and (35) in Eq.(31) we obtain
\begin{equation}
{\bf B}=\left[\frac{{\bf R}}{R}\times q\,\frac{
-R\mbox{\boldmath$\beta$} (1-\beta^2+
{\bf
R}\dot{\mbox{\boldmath$\beta$}}/c)-(R\dot{\mbox{\boldmath$\beta$}}/c)(R-
{\bf R}\mbox{\boldmath$\beta$})}{(R-{\bf R}\mbox{\boldmath$\beta$})^3}\right].
\end{equation}
If we add  ${\bf R}(1-\beta^2+ {\bf
R}\dot{\mbox{\boldmath$\beta$}}/c)$ to the numerator of the second term of
the vectorial product (38)$^{11}$ we obtain Eq.(2) (comparing with
Eq.(36)).

$$$$

{\bf ACKNOWLEDGMENTS}
We are grateful to Prof. V. Dvoeglazov and Dr. D.W.Ahluwalia for many
stimulating discussions.  We acknowledge paper of Professor E.Comay,
which put an idea into us to make present work.

\bigskip
{\small
$$$$
$^1$L.D.Landau and
E.M.Lifshitz, {\it Teoria Polia} (Nauka, Moscow, 1973) [English
translation:  {\it The Classical Theory of Field} (Pergamon, Oxford,
1975), pp. 158-160].\\
$^2$E.Comay, ``Decomposition of electromagnetic
fields into radiation and bound components", Am.J.Phys.{\bf 65},
862-867(1997).  [See p.863].\\
$^3$C.Teitelboim, D.Villarroel, and
Ch.G.van Weert, ``Classical Electrodynamics of Retarded Fields and Point
Charges", Riv.  Nuovo Cimento {\bf 3}, 1-64(1980). [See (3.25) on p.13].\\
$^4$One can calculate $\partial t_0/\partial t$  and $\partial
t_0/\partial x_i$ following Ref.1, p. 159:
$$
\frac{\partial R}{\partial t}=
\frac{\partial R}{\partial t_0}\frac{\partial t_0}{\partial t}=
-\frac{{\bf RV}}{R}\frac{\partial t_0}{\partial t}=
c\left(1-\frac{\partial t_0}{\partial t}\right),
$$
and
$$
\nabla t_0=-\frac{1}{c}\nabla R(t_0)=-\frac{1}{c}\left(
\frac{\partial R}{\partial t_0}\nabla t_0+\frac{{\bf R}}{R}\right).
$$
As a result one obtains
$$
\frac{\partial t_0}{\partial t}=\frac{1}{1-\frac{{\bf RV}}{Rc}}\qquad
{\rm and}\qquad
\frac{\partial t_0}{\partial x_i}=-
\frac{x_i-x_{0i}}{c\left(R-\frac{{\bf RV}}{c}\right)}.
$$
$^5$This follows from expressions $R=c(t-t_0)$ and ${\bf R}={\bf r}-{\bf
r}_0(t_0)$. See e.g. I.V.Saveliev, {\it Foundation of Theoretical Physics
(Osnovy Teoreticheskoi Fiziki)} (Nauka, Moscow 1975), Vol 1, ch. XIV,
{\S}78, p.  278 (in Russian). A detailed derivation of the formulas (1) and
(2) can be found in this book or the Appendix of the present paper.
We have found an interesting recent work by A.Gupta and
T.Padmanabhan ``Radiation from a charged particle and radiation reaction -
revisited" where the authors have obtained the formulas (1) and (2) by
solving Maxwell's equations in the rest frame of the charged particle
(which is a noninertial frame) and transforming the results to the
inertial frame (see hep-physics/9710036).\\
$^6$The
expressions (22)-(24) were calculated using the program ``Mathematica,
Version 2.2", therefore  it is easy to check these
calculations.\\
$^7$ D.W.Ahluwalia,``A New Type of Massive Spin-One Boson:
And Its Relation with Maxwell Equations" in {\it The Present Status of the
Quantum Theory of Light}, Eds. S.Jeffers at al (Kluwer, 1997), p.p.
443-457. A reader can also find similar speculations in the following
works:\\
S.Weinberg, ``Feynmann Rules for any Spin. II Massless
Particles", Phys. Rev.B {\bf 134} (1964), p.p. 882-896. [see p.B888, the
first statement after Eqs.  (4.21) and (4.22)];\\
D.W.Ahluwalia and
D.J.Ernst, ``Paradoxical Kinematic Acausality in Weinberg's Equations  for
Massless Particles of arbitrary Spin", Modern Phys. Lett. A {\bf 7}
(1992), p.p. 1967-1974;\\
V.V.Dvoeglazov, ``Use of the Quasipotential
Method for Describing the Vector Particle Interactions", Russian Phys. J.,
{\bf 37}, N 9 (1994), p.p. 898-902;\\
V.V.Dvoeglazov, ``Can the $2(2j+1)$
Components Weinberg-Tucher-Hammer Equations Describe the Electromagnetic
Field?", Preprint hep-th/9410174, Zacatecas, Oct. 1994.\\
$^8$In Eq.(31)
we used a well-known formula of vectorial analysis:  $$ [\nabla\times{\bf
f}]=\left[\nabla\xi\times\frac{\partial{\bf f}}{\partial \xi}\right] $$
where ${\bf f}={\bf f}(\xi)$ and $\xi=\xi(x,y,z)$.\\
$^9$In Eq.(21) $(x-x_0)V={\bf RV}$ in general.\\
$^{10}[{\bf a}\times[{\bf b}\times{\bf c}]]={\bf b}({\bf a\cdot c})-{\bf
c}({\bf a\cdot b})$.\\
$^{11}$The meaning of Eq.(38) does not change because of $[{\bf
R}\times{\bf R}]=0$.

}

 \end{document}